\documentstyle[12pt]{article}
\newcommand {\pl}{\partial}

\newcommand {\vp}{\varphi}

\newcommand {\al}{\alpha}
\newcommand {\be}{\beta}
\newcommand {\ga}{\gamma}
\newcommand {\Ga}{\Gamma}

\newcommand {\la}{\lambda}

\newcommand {\na}{\nabla}
\newcommand {\del}  {\delta}
\newcommand {\Del}  {\Delta}

\newcommand {\half}{ {\frac{1}{2}} }
\newcommand {\fourth} {\frac{1}{4} }
\newcommand {\sqg} {\sqrt{g}}

\newcommand {\Dcal}{{\cal D}}

\newcommand {\PLinv}  {{\frac{1}{\hbar}}}
\newcommand {\intx} {{\int d^2x}}

\newcommand {\ra} {\rightarrow}
\newcommand {\pr}   {{\quad .}}
\newcommand {\com}  {{\quad ,}}
\newcommand {\q}    {\quad}

\newcommand {\nn}    {\nonumber}
\newcommand {\ltsim}    {\stackrel{<}{\sim}}
\newcommand {\vs}[1]  { \vspace*{#1 cm} }

\newcounter{eq}
\newcounter{sc}

\newcommand {\MPL}  {Mod.Phys.Lett.}
\newcommand {\NP}   {Nucl.Phys.}
\newcommand {\PL}   {Phys.Lett.}

\newcommand {\CMP}  {Commun.Math.Phys.}

\begin{document}
\title{    MINBU Distribution of Two Dimensional
           Quantum Gravity\ :\ Simulation Result and Semiclassical Analysis
           \thanks{SU-95-05}
                                 }
\author{
          S. ICHINOSE
          \thanks{ E-mail address:\ ichinose@u-shizuoka-ken.ac.jp}\\
          Department of Physics, Universuty of Shizuoka,      \\
          Yada 52-1, Shizuoka 422, Japan                             \\
          N. TSUDA
          \thanks{ E-mail address:\ ntsuda@theory.kek.jp }
          and T. YUKAWA
          \thanks{ E-mail address:\ yukawa@theory.kek.jp }              \\
          National Laboratory for High Energy Physics(KEK), \\
          Tsukuba,Ibaraki 305 ,Japan                                  \\
                          }
\date{  May, 1995 }
\maketitle
\begin{abstract}
We analyse
MINBU distribution of
2 dimensional quantum gravity. New data of R$^2$-gravity  by the Monte Carlo
simulation and
its theoretical analysis
by the  semiclassical approach are presented.
The cross-over phenomenon takes place at some size of the baby universe
where the randomness competes with the smoothing force of $R^2$-term.
The dependence on the central charge $c_m$\
and on the $R^2$-coupling are
explained for the ordinary 2d quantum gravity and for $R^2$-gravity.
The $R^2$-Liouville solution plays the central role in the semiclassical
analysis.
A total derivative term (surface term) and the infrared regularization
play important roles .
The surface topology is that of a sphere .
\end{abstract}
\section{Introduction}

The quantum effects of the 2 dimensional (2d) gravitational theories are
recently measured numerically
in the computer simulation with  high statistics.
In particular the data for the entropy exponent
(string susceptibility) in 2d
quantum gravity(QG) is the same as the known exact result within
a relative precision of $O(10^{-3})$. It is due to the developement of the
simulation technique in the dynamical triangulation\cite{ADF,D,KKM}
and the findings of new observables in QG such as MINBU distribution
\cite{JM,AJT,Th}.

The data analysis is done by a rather orthodox approach,i.e., the semiclassical
approximation. It has recently been applied to 2d $R^2$-gravity and
the simulation data of $<\intx\sqg R^2>$\ and its cross-over phenomenon
are successfully explained\cite{ITY}.
We list the merits of this approach.
\begin{enumerate}
\item
The semiclassical treatment
is, at present, the unique field-theoretical approach which can analyse
the mysterious region $(25\geq )c_m\geq 1$. The conformal field theory
gives a meaningful result only for some limitted regions of $c_m$.
The Matrix model is in the similar situation.
\item
Comparison with the ordinary quantization is transparent because
the ordinary renormalizable field theories ,such as QED and QCD,
are quantized essentially in the semiclassical way.
In particular,the  renormalization properties of (2d) QG are expected to be
clarified in the semiclassical approch\cite{S1}.
\item
This approach can be used for the higher-dimensional QG such as
3d and 4d QG.
\end{enumerate}
The approach is  perturbative, therefore  choosing
the most appropriate vacuum under the global constraints (such as the area
constraint and the topology constraint)
is crucial in the proper evaluation. We explain it in Sect.3.

We add $R^2$-term to the ordinary 2d gravity
for the following reasons.
( We call the ordinary 2d gravity {\it Liouville gravity}
in contrast with {\it $R^2$-gravity} for the added one. )
\begin{enumerate}
\item
For the positive coupling,
the term plays the role of suppressing the high curvature and making
the surface smooth. For the negative one, the high curvature is energetically
favoured and making the surface rough. Therefore we can expect a richer
phase structure of the surface configuration.
\item
The term is higher-derivative ($\pl^4$), therefore it regularizes the
ultra-violet behaviour so good\cite{KPZ}. In fact the theory is
renormalizable\cite{S1}.
\item
The Einstein term ($R$-term) is topological in 2 dimension. It does not
have a local mode. The simplest interaction which is purely geometrical
and has local modes is $R^2$-term.
\item
In the lattice gravity, $R^2$-term is considered as one of natural irrelevent
terms in the continuous limit\cite{BK}.
\end{enumerate}
The $<\intx\sqg R^2>$\ simulation data
for $R^2$-gravity was presented by
\cite{TY} and the  cross-over phenomenon was clearly found.
We present here MINBU
distribution data.

\section{Lattice Simulation of 2D Quantum R$^2$-Gravity
and MINBU Distribution}
The distribution of baby universe (BU) is one of important
observables in the lattice gravity\cite{JM,AJT,Th}. It was originally
introduced
to measure the entropy exponent (string susceptibility) efficiently.
Fig.1 shows the configuration of a BU with an area B (variable) from the mother
universe with an area A (fixed).

{\vs 6}
\begin{center}
Fig.1\q MINBU configuration
\end{center}
The 'neck' of Fig.1 is composed of three links which is
the minimum loop in the dynamically triangulated surface. The configuration
is called
the minimum neck baby universe (MINBU). MINBU distribution for the
Liouville-gravity and its matter-coupled case
were already measured\cite{AJT,Th,AT}.

First we explain briefly our lattice model of $R^2$-gravity.
The surface is regularized by the triangulation. The number of vertices
,where some links (edges of triangles) meet, is $N_0$. The number of
links at the i-th vertex ($i=1,2\cdots,N_0$) is $q_i$.
The number of triangles($N_2$) is related to $N_0$\ as
$N_2=2N_0-4$\ for the sphere topology.
The discretized model is then
described by
\begin{eqnarray}
&S_L=-\be_L\frac{4\pi^2}{3}\sum_{i=0}^{N_0}\frac{(6-q_i)^2}{q_i}
=-48\pi^2\be_L\sum_{i}\frac{1}{q_i}+\mbox{const}\com &\label{lat.1}
\end{eqnarray}
where $\be_L$\ is the R$^2$-coupling constant of the lattice model.
We do measurement for
$\be_L=0,50,100,200,300,-20,-50$\ .
We present the MINBU dstribution of $R^2$-gravity with no matter field
(pure R$^2$-gravity) in Fig.2 and 3 for
$\be_L\geq 0$\ and for $\be_L\leq 0$ respectively.
The total number of triangles
is $N_2=5000$.
For the detail see \cite{TY}.

{\vs 6}
\begin{center}
Fig.2\q MINBU distribution for $\be_L\geq 0$, Pure $R^2$-gravity.
\end{center}

{\vs 6}
\begin{center}
Fig.3\q MINBU distribution for $\be_L\leq 0$. Pure $R^2$-gravity.
\end{center}

As for positive $\be_L$\ (Fig.2), we see clearly the transition point $P_0$
, for each curve,
at which the distribution qualitatively changes. For the region $P=B/A\ > P_0$,
the birth probability
decreases as the size of BU increases.
For the region $P < P_0$,
the birth probability increases as the size
of BU increases.
The value of the transition point $P_0$\ depends on $\be$ \ and increases
as $\be$\ increases.
As for negative $\be$\ (Fig.3), the slope of the curve
tends to be sharp as $|\be|$\ increases at least for the region $P<P_1$.
The transition point $P_1$\ is not so clear as Fig.2.

In Sect 4.2 we interpret these data theoretically using the
semiclassical approach explained in Sect 3.

\section{Semiclassical Approach}
We analyse the simulation data by the semiclassical approach. The $R^2$-
gravity interacting with $c_m$-components scalar matter fields is
described by
\begin{eqnarray}
& S=\intx\sqg (\frac{1}{G} R-\be R^2-\mu
-\half\sum_{i=1}^{c_m}\pl_a\Phi_i\cdot
g^{ab}\cdot \pl_b\Phi_i)\com\q (\ a,b=1,2\ )\com            & \label{3.1}
\end{eqnarray}
where
$G$\ is the gravitaional coupling constant, $\mu$\ is the cosmological
constant ,
$\be$\ is the coupling strength for $R^2$-term and $\Phi$\ is the $c_m$-
components scalar matter fields. The signature is Euclidean.
The partition function ,
under the fixed area condition\
$ A=\intx \sqg\ $ and with the conformal-flat gauge\
$g_{ab}=\ e^{\vp}\ \del_{ab}$\ , is written as
\cite{P},
\begin{eqnarray}
& {\bar Z}[A]=
\int\frac{\Dcal g\Dcal\Phi}{V_{GC}}\{exp\PLinv S\}~\del(\intx\sqg-A)
=exp\PLinv (\frac{8\pi(1-h)}{G}-\mu A)\times Z[A]\com & \nn\\
& Z[A]\equiv\int\Dcal\vp~ e^{+\frac{1}{\hbar}
S_0[\vp]}~\del(\intx ~e^\vp - A)\com &   \label{3.2}\\
& S_0[\vp]=\intx\ (\frac{1}{2\ga}\vp\pl^2\vp
-\be~e^{-\vp}(\pl^2\vp)^2 +\frac{\xi}{2\ga}\pl_a(\vp\pl_a\vp)\ )\com
\q \frac{1}{\ga}=\frac{1}{48\pi}(26-c_m)\com & \label{3.3}
\end{eqnarray}
where  $h$\ is the number of handles
\footnote{
The sign for the action is different from the usual convention as seen in
(\ref{3.2}).
}.
$V_{GC}$\ is the gauge volume due to the general coordinate invariance.
$\xi$\ is a free parameter. The total derivative term generally appears when
integrating out the anomaly equation
\ $\del S_{ind}[\vp]/\del\vp=\frac{1}{\ga}\pl^2\vp\ $.
This term turns out to be very important.
\footnote{
The uniqueness of this term, among all possible total derivatives, is shown
in \cite{ITY}.
}
We consider the manifold of a fixed topology of
the sphere ,$h=0$\ and  the case $\ga>0\ (c_m<26)$.
\ $\hbar$\ is  Planck constant.
\footnote{
In this section only,we explicitly write $\hbar$\ (Planck constant) in order
to show the perturbation structure clearly.
}

$Z[A]$\ is rewritten as, after the Laplace transformation and the inverse
Laplace one,
\begin{eqnarray}
& Z[A]=\int\frac{d\la}{\hbar}\int\Dcal\vp~exp\ \frac{1}{\hbar}[\ S_0[\vp]
-\la (\intx e^\vp - A)]                                    &\nn\\
&=\int\frac{d\la}{\hbar}e^{\PLinv \la A}
\int\Dcal\vp~exp~\{\PLinv S_\la[\vp]\}\com &\nn\\
&S_\la[\vp]\equiv S_0[\vp]-\la\intx~e^\vp\ &\nn\\
&=\intx\ (\frac{1}{2\ga}\vp\pl^2\vp -\be~e^{-\vp}(\pl^2\vp)^2\
+\frac{\xi}{2\ga}\pl_a(\vp\pl_a\vp)\
 -\la~e^\vp\ )\com    &\label{3.4}
\end{eqnarray}
where the $\la$-integral should be carried out along an appropriate
contour parallel to the imaginary axis in the complex $\la$-plane.
Note that the $\del$-function
constraint in (\ref{3.2}) is substituted by the $\la$-integral.
The leading order configuration is given by the stationary minimum.
\begin{eqnarray}
\left. \frac{\del S_\la[\vp]}{\del\vp}\right|_{\vp_c}
=\left. \frac{1}{\ga}\pl^2\vp
+\be\{ e^{-\vp}(\pl^2\vp)^2-2\pl^2(e^{-\vp}\pl^2\vp)\}-\la e^\vp
                                               \right|_{\vp_c}=0\com\nn\\
\left.\frac{d}{d\la}(\la A+S_\la[\vp_c])\right|_{\la_c}=0\com
                                                         \label{3.5}\\
Z[A]\approx \PLinv exp~\PLinv\{\la_cA+S_{\la_c}[\vp_c]\}\equiv
\PLinv exp~\PLinv \Ga^{eff}_c \pr\nn
\end{eqnarray}
Generally this approximation is valid for a large system.
In the present case, the system size is proportional to
$\frac{4\pi}{\ga}=\frac{26-c_m}{12}$. We expect the approximation
is valid except the region: $c_m\sim 26$.

The solution $\vp_c$\ and $\la_c$\ ,which describes the positive-constant
curvature solution and is continuous at $\be=0$,\ are given by\cite{ITY}
\begin{eqnarray}
\vp_c(r )=-ln~\{ \frac{\al_c}{8}(1+\frac{r^2}{A})^2\}\com\q
r^2=(x^1)^2+(x^2)^2\com                            \nn\\
\al_c=\frac{4\pi}{w}\{ w+1-\sqrt{w^2+1 -2\xi w} ~\}\com\q
w=16\pi\be'\ga\com\q \be'\equiv \frac{\be}{A}\com                 \label{3.6}\\
\ga\la_c A=\frac{w}{16\pi}(\al_c)^2-\al_c\com\nn
\end{eqnarray}
where $\xi$\ must satisfy
$-1\leq\ \xi\ \leq\ +1$\ for the realness of $\al_c$.
$(x^1,x^2)$\ are the flat (plane) coordinates.
The partition function at the classical level
is given by
\begin{eqnarray}
& \Ga^{eff}_c=~ln~Z[A]|_{\hbar^0}
=\la_c A+(1+\xi)\frac{4\pi}{\ga}~ln\frac{\al_c}{8}-\frac{\al_c}{\ga}w
+C(A)\com &                                          \nn\\
& C(A)=\frac{8\pi (2+\xi)}{\ga}+\frac{8\pi\xi}{\ga}
\{~ln(L^2/A)-1~\}+O(A/L^2)\com      &
                                             \label{3.7}\\
& \frac{L^2}{A}\gg 1\com &\nn
\end{eqnarray}
where
$L$\ is the {\it infrared cut-off}
($r^2\leq L^2$)
introduced for the divergent volume integral of the total derivative term.
Note that $C(A)$\ does not depend on $\be$\ (or $w$) but on $c_m$\ (or $\ga$)
and $A$. Furthermore $C(A)$\ has an arbitrary constant of the form
$(8\pi\xi/\ga)\times\mbox{(const)}$\ due to the freedom of the choice of
the regularization parameter:\ $L\ra \mbox{(const)}'\times L$.
This arbitrary constant turns out to be important.

For the case $\be=0$\ , the theory is ordinary 2d gravity and we call it
Liouville gravity in contrast with $R^2$-gravity for $\be\not= 0$.
For the case $c_m=0$\ , the theory is called the pure gravity in contrast
with the matter-coupled gravity $c_m\not= 0$.

\section{Semiclassical Analysis of MINBU Distribution}

First we explain the free parameter $\xi$.
Recent analysis of the present theory at the (1-loop) quantum level has
revealed that it is conformal (the renormalization group beta functions=0)
for $w\geq 1$\ when we take $\xi=1$\ \cite{S1}.
Therefore the value $\xi=1$\ has some meaning purely within the theory.
The validity of this choice is also confirmed from a different approach,
that is,
the comparison of
the special case $\be$(or $w$)$=0$\ (Liouville gravity) of the present result
with the corresponding
result from the conformal field theory (KPZ result)\cite{KPZ}.
The asymptotic behaviour of $Z[A]|_{\hbar^0}$ \ at $w=0$\
is given, from (\ref{3.7}), as
\begin{eqnarray}
Z[A]|_{\hbar^0,w=0}=\left.e^{\Ga^{eff}_c}\right|_{w=0}
= exp\{ \frac{4\pi}{\ga}(3-\xi)
+(1+\xi)\frac{4\pi}{\ga}ln\frac{1+\xi}{2}+\frac{8\pi\xi}{\ga}ln\frac{L^2}{A}
\}\approx                                             \nn\\
A^{-\frac{8\pi\xi}{\ga}}\times\mbox{const}=A^{-\frac{26-c_m}{6}\xi}
\times\mbox{const}\com\nn\\
\mbox{as}\ A\rightarrow +\infty\pr\label{cdep.1}
\end{eqnarray}
On the other hand, the KPZ result is
\begin{eqnarray}
Z^{KPZ}[A]\sim A^{\ga_s-3}\com\
\ga_s=\frac{1}{12}\{c_m-25-\sqrt{(25-c_m)(1-c_m)}\}+2\pr \label{cdep.2}
\end{eqnarray}
In order for our result to coincide with the KPZ result in the 'classical
limit' $c_m\rightarrow -\infty$\ :\
$Z^{KPZ}[A]\sim A^{+\frac{1}{6}c_m}$\ , we must take
\begin{eqnarray}
\xi=1\com                \label{cdep.3}
\end{eqnarray}
in (\ref{cdep.1}).
In the following of this text we take this value.
\footnote{
In the numerical evaluation, we take $\xi=0.99$~ for the practical reason.
}

The asymptotic behaviour of
the present semiclassical result for the Liouville gravity
is, taking $\xi=1$\ in (\ref{cdep.1}),
\begin{eqnarray}
Z[A]\sim A^{-\frac{26-c_m}{6}}\times A^{-1}\com\q A\rightarrow +\infty
\com  \label{cdep.4}
\end{eqnarray}
where the additional factor $A^{-1}$~ comes from the $\la$-integral in
the expression of $Z[A]$, (\ref{3.4})\cite{S2}.
Now we compare the KPZ result and the semiclassical result in the
normalized form.
\begin{eqnarray}
Z^{KPZ}_{norm}[A]\equiv \frac{Z^{KPZ}[A]}{Z^{KPZ}[A]|_{c_m=0}}
\sim A^{\ga_s(c_m)-\ga_s(c_m=0)}\com\nn\\
\ga_s(c_m)-\ga_s(c_m=0)=\frac{1}{12}\{c_m+5-\sqrt{(25-c_m)(1-c_m)}\}
                                            \com   \label{cdep.4b}\\
Z_{norm}[A]\equiv \frac{Z[A]}{Z[A]|_{c_m=0}}
\sim A^{+\frac{c_m}{6}} \pr  \nn
\end{eqnarray}
We can numerically confirm that the semiclassical result, $\frac{c_m}{6}$,
and the KPZ result, $\ga_s(c_m)-\ga_s(c_m=0)$~, have very similar behaviour
for the region $c_m\leq 1$\cite{S2}.

Now we go back to the general value of $\be$.
The birth-probability
of the baby universe
with area $B (0 < B < A/2)$ from the mother universe with the total area A
is given by\cite{JM}
\begin{eqnarray}
{n_A(B)}=\frac
{3(A-B+a^2)(B+a^2)Z[B+a^2]Z[A-B+a^2]}
{A^2\times Z[A]}                                 \nn\\
\approx\frac
{3(1-p)pZ[pA]Z[(1-p)A]}{Z[A]}\com            \label{cdep.5}\\
\ln~(\frac{ {n_A(B)} }{3})
\approx\ln~(1-p)p+\ln~Z[pA]+\ln~Z[(1-p)A]-\ln~Z[A]\com\nn\\
p\equiv\frac{B}{A}\com\q 0<p<\half\pr         \nn
\end{eqnarray}
We apply the  result of $Z[A]$\ in Sect.3
to the above expressions.

\subsection{ $c_m$-dependence}

First we present the semiclassical prediction for Liouville gravity($\be=0$).
The result  (\ref{3.7}) for the case $\be=0$\ gives ,taking $\xi=1$,
\begin{eqnarray}
 \ga~ln~Z[rA]
=8\pi (\ln~\pi+1)+8\pi\ln(\frac{1}{r}\cdot \frac{L^2}{A})\pr
                                             \label{cdep.6}
\end{eqnarray}
Then the MINBU distribution normalized by the pure garvity ($c_m=0$)
is obtained as
\begin{eqnarray}
\frac{n_A(B)}{n_A(B)|_{c_m=0}}=
 \{p(1-p)\}^{\frac{c_m}{6}}\times
\exp~\{ \frac{c_m}{12}\times \Del\}\com\nn\\
\Del\equiv -2(\ln~\pi +1)-2\ln\frac{L^2}{A}\com
                                             \label{cdep.7}
\end{eqnarray}
where $\Del$\ can be regarded as the free real parameter due to
the arbitrariness of
the infrared regularization parameter $L$.
We know from the result (\ref{cdep.7}) that the MINBU distribution lines
for different $c_m$'s cross at the single point $p=p^{*}$\ given by
\begin{eqnarray}
p^*(1-p^*)=\exp\{-\half~\Del\}\com\q p^*<\half\pr
                                             \label{cdep.8}
\end{eqnarray}
Fig.4 shows three typical cases of $p^*$\ .

\vspace{2cm}

{\vs 6}
\begin{center}
Fig.4\q Three typical cases of the solution of (\ref{cdep.8}).
\end{center}
The choice of $\Del$\ is important to fit the theoretical curve (\ref{cdep.7})
with the data.
We show the behaviour of (\ref{cdep.7}) for the three
cases:\ 1)\ $\exp(-\half \Del)~\ll \fourth$\ ,Near Point O,Fig.5a\ ;\
2)\ $\exp(-\half \Del)~>\fourth$\ ,Above Point A ,Fig.5b\ ;\
3)\ $\exp(-\half \Del)~=\fourth -0$\ ,Near Point A ,Fig.5c.

{\vs 6}
\begin{center}
Fig.5a\q MINBU distribution for Liouville gravity, $\Del=8$
\end{center}

{\vs 6}
\begin{center}
Fig.5b\q MINBU distribution for Liouville gravity, $\Del=1$
\end{center}

{\vs 6}
\begin{center}
Fig.5c\q MINBU distribution for Liouville gravity, $\Del=3$
\end{center}

Fig.5a well fits with the known result of the computer
simulation\cite{AJT,AT}.
This result shows the importance of the infrared regularization.

\subsection{$\be$-dependence}
We consider the pure gravity($c_m=0$).
We  plot MINBU dstribution, $ln~n_A(B)$,
as the function of $p\ (\ 0.001<p<0.1\ )$\
for various cases of $\be'=\be/A$\
($\xi=0.99$).
Fig.6a and 6b show that for $\be'>0$\ and $\be'<0$\ respectively.

{\vs 5}
\begin{center}
Fig.6a\q MINBU distribution for $\be'\geq 0$. $\xi=0.99,c_m=0$.
\end{center}

{\vs 5}
\begin{center}
Fig.6b\q MINBU distribution for $\be'\leq 0$. $\xi=0.99,c_m=0$.
\end{center}

The above results of Fig.6a and Fig.6b qualitatively coincide with
those of Fig.2 and Fig.3, respectively.

\vspace{1cm}

We list the asymptotic behaviour of $\ln~n_A(B)$\
for the general $\xi$\ and $c_m$\ in Table 1.

\vspace{0.5cm}
\begin{tabular}{|c|c|c|c|}
\hline
Phase         & (C)\ $0<p\ll -w(\ltsim 1)$
                & (B)\ $|w|\ll p$           & (A)\ $0<p\ll w(\ltsim 1)$    \\
\hline
 $\al^-_p(pA)$ & $8\pi\{1+\frac{1-\xi}{2}\frac{p}{w}$
                 &  $4\pi(1+\xi)\{1-\frac{1-\xi}{2}\frac{w}{p}$
                                      & $\frac{4\pi(1+\xi)p}{w}\times $  \\
    & $+O(\frac{p^2}{w^2})\}$    & $+O(\frac{w^2}{p^2})\}$
                                      & $\{1+O(\frac{p}{w})\}$          \\
\hline
    & $(1-\frac{8\pi\xi}{\ga})\ln~p$  & $(1-\frac{8\pi\xi}{\ga})\ln~p$
                                & $\{1-\frac{4\pi(1-\xi)}{\ga}\}\ln~p$   \\
 $\ln~{n_A(B)}$   &$-\frac{4\pi}{\ga}\frac{w}{p}$
                      &   $+O(\frac{w}{p})$
                             & $-\frac{4\pi(1+\xi)}{\ga}~\ln~w$ \\
  &$+O(\frac{p}{w})$ & +SmallTerm
            & $+O(\frac{p}{w}) $              \\
& +SmallTerm &
                         & +SmallTerm        \\
\hline
\multicolumn{4}{c}{\q}                                   \\
\multicolumn{4}{c}{Table 1\ \  Asymp. behaviour of
MINBU distribution,
(\ref{cdep.5}), }\\
\multicolumn{4}{c}{ for general $c_m$~ and $\xi$.
$R>0, w\equiv 16\pi\be'\ga, \ga=\frac{48\pi}{26-c_m}>0,p=\frac{B}{A},$}\\
\multicolumn{4}{c}{ $0<p\ll 1,\ |w|\ltsim 1,\ \mbox{SmallTerm}= \mbox{const}
+O(wp)+O(p).$}
\end{tabular}
\vspace{0.5cm}

We characterize each phase in Table 1  as follows.
\flushleft{(A)\ $0<p\ll w$:\ Smoothly Creased Surface
\footnote{
In \cite{ITY} we called it Free Creased Surface because this is the phase
where the free kinetic term ($R^2$-term) dominates.}
          }

The smoothing term, $R^2$, dominates the main configuration and the surface
is smooth. The left part $P<P_0(w)$\ for each curve ($w$) in Fig.6a corresponds
to this phase.
The small BU is harder to be born because it needs high-curvature
locally. The large BU is energetically preferable to be born.
The area constraint is not effective in this phase.
The characteristic scale is $\be$.
\flushleft{(B)\ $|w|\ll p$:\ Fractal Surface}

The randomness dominates the configuration.
The size of BU is so enough large that the $R^2$-term is not effective.
The area constraint is neither effective. There is no characteristic scale.
The right part $P>P_0(w)$\ for each
curve ($w$) in Fig.6a and the right part $P>P_1(w)$\ for each curve ($w$) in
Fig.6b correspond to this phase.
The MINBU distribution is mainly determined
by the random distribution of the surface configuration\cite{BIPZ}.
\flushleft{(C)\ $0<p\ll -w$:\ Rough Surface
\footnote{
In \cite{ITY} we called it Strongly Tensed Perfect Sphere because the surface
tension is negatively large and the shape of the whole surface is near
a sphere. At the same time the surface tend to become
sharp-pointed  because it increases the curvature.
We call the surface under this circumstace,simply, Rough Surface.}
          }

Due to the large negative value of $R^2$-coupling, the configuration
with the large curvature is energetically preferable on the one hand,
it is strongly influenced by the area constraint on the other hand.
Therefore the large BU is much harder to be born than (B) because
it has a small curvature and a large area.
The small
BU is much easier to be born than (B) because it has a large curvature
and a small area. The left part $P<P_1(w)$\ for each curve ($w$) in Fig.6b
corresponds to this phase. The characteristic scale is the total area $A$.

\vspace{1cm}
\q We see the phase structure of Table 1 is the same as that of \cite{ITY}
by the substitution of $w$\ by $w/p$\ .
Although both simulations measure the same surface property,
the cross-over phenomenon,however, appears  differently.
In \cite{ITY} the physical quantity
$<\intx\sqg R^2>$\ is taken to see the surface property.
The cross-over can be seen only by measuring for a range
of $w$\ and the transition point is given by a certain value
$|w^*|\approx 1$.
This is contrasting with the present case.
The cross-over can be seen
for any $w$. The transition  is seen at the point $p^*$\ ,in the
MINBU distribution,
given by $|w|/p^*\approx 1$.
We understand as follows.
The MINBU distribution measures the surface
at many different 'scales' $B$, whereas the quantity $<\intx\sqg R^2>$\
measures the surface at a fixed 'scale'( $B_1$\ (or $p_1$) in the MINBU
terminology).

\subsection{ General Case}
We consider the general case of $c_m$\ and $\be$. This general case
is not yet measured by the Monte Carlo simulation. We present
the semiclassical prediction.
The analysis so far shows
the normalization
((\ref{cdep.4b}) and (\ref{cdep.7}))
and the choice of an arbitrary
constant due to the infrared regularization (\ref{cdep.7})
are important
for the quantitative adjustment.
Here, however, we are content with the qualitative behaviour.
We donot do the normalization and we ignore the $\ln~\frac{L^2}{A}$~term
in the  evaluation of this subsection.

\flushleft{(1)\ $c_m$-dependence}

We  stereographically show
MINBU distributions for the
range:\ $0.001\leq p\leq 0.2,\ -24\leq c_m\leq +24$\ ,
in Fig.7a($\be'=0$) , Fig.7b($\be'=+10^{-4}$) and
Fig.7c($\be'=-10^{-5}$).

{\vs 5}
\begin{center}
Fig.7a\q MINBU dstribution for $0.001\leq p\leq 0.2,\ -30\leq c_m\leq +24$\ .
$\be'=0,\xi=0.99$.
\end{center}

{\vs 5}
\begin{center}
Fig.7b\q MINBU dstribution for $0.001\leq p\leq 0.2,\ -30\leq c_m\leq +24$\ .
$\be'=+10^{-4},\xi=0.99$.
\end{center}

{\vs 5}
\begin{center}
Fig.7c\q MINBU dstribution for $0.001\leq p\leq 0.2,\ -30\leq c_m\leq +24$\ .
$\be'=-10^{-5},\xi=0.99$.
\end{center}

No 'ridge' appears in Fig.7a.
{}From this, we see matter fields
affect the surface dynamics homogeneously at all scales.
(This result is natural because the matter coupling constand $c_m$\ does
not have the scale dimension.) The slope along the $p$-axis continuously
decreases as $c_m$\ increases.
In Fig.7b,
a ridge runs from a low $p$\ to a high $p$\ as $c_m$\ increases.
In Fig.7c,a 'hollow'  runs from a high $p$\
to a low $p$\ as $c_m$\ increases.
The ridge and the hollow correspond to the series of the cross-over points.
In both Fig.7b and Fig.7c, the cross-over
becomes dimmer as $c_m$\ increases and becomes sharper as $c_m$\ decreases.

\flushleft{(2)\ $\be$-dependence}

We  stereographically show
MINBU distributions for the
range:\ $0.001\leq p\leq 0.2,\ -10^{-5}\leq \be'\leq +10^{-4}$\ ,
in Fig.8a($c_m=0$) , Fig.8b($c_m=+10$) and
Fig.8c($c_m=-10$).

{\vs 5}
\begin{center}
Fig.8a\q MINBU dstribution for $0.001\leq p\leq 0.2,
\ -10^{-5}\leq \be'\leq +10^{-4}$\ .
$c_m=0,\xi=0.99$.
\end{center}

{\vs 5}
\begin{center}
Fig.8b\q MINBU dstribution for $0.001\leq p\leq 0.2,
\ -10^{-5}\leq \be'\leq +10^{-4}$\ .
$c_m=+10,\xi=0.99$.
\end{center}

{\vs 5}
\begin{center}
Fig.8c\q MINBU dstribution for $0.001\leq p\leq 0.2,
\ -10^{-5}\leq \be'\leq +10^{-4}$\ .
$c_m=-10,\xi=0.99$.
\end{center}

The Fig.8a corresponds to the stereographic display of Fig.6a and 6b.
In each of Fig.8a-c, a ridge appears for $\be'>0$\ . For $\be'<0$\ ,
a tower appears instead of a ridge. For a large positive $c_m$\
( matter dominated region, $c_m=10$\ in Fig.8b) the undulation
of the MINBU dstribution surface
\footnote{Do not confuse it with the 2d manifold which
the present model of gravity represents.}
is small(the cross-over is dim), whereas it is large(the cross-over is sharp)
for a large negative $c_m$\
(matter anti-dominated region, $c_m=-10$\ in Fig.8c).

\section{Discussion and Conclusion}

In the (2d) QG,at present, there exists no simple way to find good
physical observables. They have been found by 'try and error'. MINBU is one
of  good observables to measure the surface property. Quite recently
a new observable ,the 'electric resistivity' of the surface,
is proposed by \cite{KTY}.
By measuring the observable for
the matter-coupled Liouville gravity,
they observe a cross-over ,near $c_m=1$\ ,from the surface
where a complex-structure is well-defined to
the surface where it is not well-defined.
The analysis of
the new obserbable, from the standpoint of the present approach, is important.

\q There are some straightforward but important applications of
the present analysis
:\ 1)\ higher-genus case, 2)\ the case with
other higher-derivative terms such as $R^3$\ and $\na R\cdot \na R$\ ,
3)\ the quantum effect.
As for 2) ,references \cite{TY2} and \cite{Tsuda}
have already obtained the Monte Carlo data.

\q We have presented the numerical result of MINBU and its theoretical
explanation using the semiclassical approximation.
The surface properties are characterized.
It is confirmed that the present lowest approximation
is very efficient to analyse 2d quantum gravity, at least, qualitatively.

\q Finally we expect
other new observables will be found and many Monte Carlo
measurements will be done ,including 3 and 4 dimensional cases,
next a few years.
The interplay between the measurement by the
computer simulation and the theoretical interpretation will become important
more and more.
We believe this process
will lead to the right understanding of the (Euclidean) quantum gravity.

\begin{flushleft}
{\bf Acknowledgement}
\end{flushleft}
The authors thank N. Ishibashi and H. Kawai for comments and discussions
about the present work.


\newpage
\begin{flushleft}
{\bf Figure Captions}
\end{flushleft}

\begin{itemize}
\item
Fig.1\q MINBU configuration.
\item
Fig.2\q MINBU distribution for $\be_L\geq 0$, Pure $R^2$-gravity.
\item
Fig.3\q MINBU distribution for $\be_L\leq 0$. Pure $R^2$-gravity.
\item
Fig.4\q Three typical cases of the solution of (\ref{cdep.8}).
\item
Fig.5a\q MINBU distribution for Liouville gravity, $\Del=8$.
\item
Fig.5b\q MINBU distribution for Liouville gravity, $\Del=1$.
\item
Fig.5c\q MINBU distribution for Liouville gravity, $\Del=3$.
\item
Fig.6a\q MINBU distribution for $\be'\geq 0$. $\xi=0.99,c_m=0$.
\item
Fig.6b\q MINBU distribution for $\be'\leq 0$. $\xi=0.99,c_m=0$.
\item
Fig.7a\q MINBU dstribution for $0.001\leq p\leq 0.2,\ -30\leq c_m\leq +24$\ .
$\be'=0,\xi=0.99$.
\item
Fig.7b\q MINBU dstribution for $0.001\leq p\leq 0.2,\ -30\leq c_m\leq +24$\ .
$\be'=+10^{-4},\xi=0.99$.
\item
Fig.7c\q MINBU dstribution for $0.001\leq p\leq 0.2,\ -30\leq c_m\leq +24$\ .
$\be'=-10^{-5},\xi=0.99$.
\item
Fig.8a\q MINBU dstribution for $0.001\leq p\leq 0.2,
\ -10^{-5}\leq \be'\leq +10^{-4}$\ .
$c_m=0,\xi=0.99$.
\item
Fig.8b\q MINBU dstribution for $0.001\leq p\leq 0.2,
\ -10^{-5}\leq \be'\leq +10^{-4}$\ .
$c_m=+10,\xi=0.99$.
\item
Fig.8c\q MINBU dstribution for $0.001\leq p\leq 0.2,
\ -10^{-5}\leq \be'\leq +10^{-4}$\ .
$c_m=-10,\xi=0.99$.
\end{itemize}

\end{document}